\documentstyle[twoside]{article} 
\input amssym.def
\catcode`\@=11
\long\def\@makefntext#1{
\protect\noindent \hbox to 3.2pt {\hskip-.9pt  
$^{{\eightrm\@thefnmark}}$\hfil}#1\hfill}               

\def\thefootnote{\fnsymbol{footnote}}
\def\@makefnmark{\hbox to 0pt{$^{\@thefnmark}$\hss}}    
        
\def\ps@myheadings{\let\@mkboth\@gobbletwo
\def\@oddhead{\hbox{}
\rightmark\hfil\eightrm\thepage}   
\def\@oddfoot{}\def\@evenhead{\eightrm\thepage\hfil
\leftmark\hbox{}}\def\@evenfoot{}
\def\sectionmark##1{}\def\subsectionmark##1{}}

\def\qed{\hbox{${\vcenter{\vbox{                        
   \hrule height 0.4pt\hbox{\vrule width 0.4pt height 6pt
   \kern5pt\vrule width 0.4pt}\hrule height 0.4pt}}}$}}

\renewcommand{\thefootnote}{\fnsymbol{footnote}}        

\def\theequation{\thesectionc.\arabic{equation}}        

\oddsidemargin=\evensidemargin
\renewcommand{\thefootnote}{\fnsymbol{footnote}}

\newcounter{sectionc}\newcounter{subsectionc}\newcounter{subsubsectionc}
\renewcommand{\section}[1] {\vspace{12pt}\addtocounter{sectionc}{1} 
\setcounter{subsectionc}{0}\setcounter{subsubsectionc}{0}\noindent 
        {\tenbf\thesectionc. #1}\par\vspace{5pt}}
\renewcommand{\subsection}[1] {\vspace{12pt}\addtocounter{subsectionc}{1} 
        \setcounter{subsubsectionc}{0}\noindent 
        {\bf\thesectionc.\thesubsectionc. {\kern1pt \bfit #1}}\par\vspace{5pt}}
\renewcommand{\subsubsection}[1] {\vspace{12pt}\addtocounter{subsubsectionc}{1}
        \noindent{\tenrm\thesectionc.\thesubsectionc.\thesubsubsectionc.
        {\kern1pt \tenit #1}}\par\vspace{5pt}}
\newcommand{\nonumsection}[1] {\vspace{12pt}\noindent{\tenbf #1}
        \par\vspace{5pt}}

\newcounter{appendixc}
\newcounter{subappendixc}[appendixc]
\newcounter{subsubappendixc}[subappendixc]
\renewcommand{\thesubappendixc}{\Alph{appendixc}.\arabic{subappendixc}}
\renewcommand{\thesubsubappendixc}
        {\Alph{appendixc}.\arabic{subappendixc}.\arabic{subsubappendixc}}

\renewcommand{\appendix}[1] {\vspace{12pt}
        \refstepcounter{appendixc}
        \setcounter{figure}{0}
        \setcounter{table}{0}
        \setcounter{lemma}{0}
        \setcounter{theorem}{0}
        \setcounter{corollary}{0}
        \setcounter{definition}{0}
        \setcounter{equation}{0}
        \setcounter{proposition}{0}
        \renewcommand{\thefigure}{\Alph{appendixc}.\arabic{figure}}
        \renewcommand{\thetable}{\Alph{appendixc}.\arabic{table}}
        \renewcommand{\theappendixc}{\Alph{appendixc}}
        \renewcommand{\thelemma}{\Alph{appendixc}.\arabic{lemma}}
        \renewcommand{\thetheorem}{\Alph{appendixc}.\arabic{theorem}}
        \renewcommand{\theproposition}{\Alph{appendixc}.\arabic{proposition}}
        \renewcommand{\thedefinition}{\Alph{appendixc}.\arabic{definition}}
        \renewcommand{\thecorollary}{\Alph{appendixc}.\arabic{corollary}}
        \renewcommand{\theequation}{\Alph{appendixc}.\arabic{equation}}
        \noindent{\tenbf Appendix#1}\par\vspace{5pt}}
\newcommand{\subappendix}[1] {\vspace{12pt}
        \refstepcounter{subappendixc}
        \noindent{\bf Appendix \thesubappendixc. {\kern1pt \bfit #1}}
        \par\vspace{5pt}}
\newcommand{\subsubappendix}[1] {\vspace{12pt}
        \refstepcounter{subsubappendixc}
        \noindent{\rm Appendix \thesubsubappendixc. {\kern1pt \tenit #1}}
        \par\vspace{5pt}}

\newcommand{\textlineskip}{\baselineskip=13pt}
\newcommand{\smalllineskip}{\baselineskip=10pt}

\def\eightcirc{
\begin{picture}(0,0)
\put(4.4,1.8){\circle{6.5}}
\end{picture}}
\def\eightcopyright{\eightcirc\kern2.7pt\hbox{\eightrm c}} 

\newcommand{\copyrightheading}[1]
        {\vspace*{-2.5cm}\smalllineskip{\flushleft
        {\footnotesize Mathematical Models and Methods in Applied Sciences #1}\\
        {\footnotesize $\eightcopyright$\, World Scientific Publishing
         Company}\\
         }}

\def\abstracts#1#2#3{{
        \centering{\begin{minipage}{4.5in}
           \baselineskip=10pt\footnotesize
        \parindent=0pt #1\par 
        \parindent=15pt #2\par
        \parindent=15pt #3
        \end{minipage}}\par}} 

\def\keywords#1{{
        \centering{\begin{minipage}{4.5in}\baselineskip=10pt\footnotesize
        {\footnotesize\it Keywords}\/: #1
         \end{minipage}}\par}}

\renewenvironment{thebibliography}[1]
        {\frenchspacing
         \ninerm\baselineskip=11pt
         \begin{list}{\arabic{enumi}.}
        {\usecounter{enumi}\setlength{\parsep}{0pt}     
         \setlength{\leftmargin 12.7pt}{\rightmargin 0pt} 
         \setlength{\itemsep}{0pt} \settowidth
        {\labelwidth}{#1.}\sloppy}}{\end{list}}

\newcounter{itemlistc}
\newcounter{romanlistc}
\newcounter{alphlistc}
\newcounter{arabiclistc}

\def\@citex[#1]#2{\if@filesw\immediate\write\@auxout
        {\string\citation{#2}}\fi
\def\@citea{}\@cite{\@for\@citeb:=#2\do
        {\@citea\def\@citea{,}\@ifundefined
        {b@\@citeb}{{\bf ?}\@warning
        {Citation `\@citeb' on page \thepage \space undefined}}
        {\csname b@\@citeb\endcsname}}}{#1}}

\newif\if@cghi
\def\cite{\@cghitrue\@ifnextchar [{\@tempswatrue
        \@citex}{\@tempswafalse\@citex[]}}
\def\citelow{\@cghifalse\@ifnextchar [{\@tempswatrue
        \@citex}{\@tempswafalse\@citex[]}}
\def\@cite#1#2{{$\null^{#1}$\if@tempswa\typeout
        {IJCGA warning: optional citation argument 
        ignored: `#2'} \fi}}
\newcommand{\citeup}{\cite}

\def\pmb#1{\setbox0=\hbox{#1}
        \kern-.025em\copy0\kern-\wd0
        \kern.05em\copy0\kern-\wd0
        \kern-.025em\raise.0433em\box0}

\def\fnm#1{$^{\mbox{\scriptsize #1}}$}
\def\fnt#1#2{\footnotetext{\kern-.3em
        {$^{\mbox{\scriptsize #1}}$}{#2}}}

\def\fpage#1{\begingroup
\voffset=.3in
\thispagestyle{empty}\begin{table}[b]\centerline{\footnotesize #1}
        \end{table}\endgroup}

\def\runninghead#1#2{\pagestyle{myheadings}
\markboth{{\protect\footnotesize\it{\quad #1}}\hfill}
{\hfill{\protect\footnotesize\it{#2\quad}}}}
\font\tenrm=cmr10
\font\tenit=cmti10 
\font\tenbf=cmbx10
\font\bfit=cmbxti10 at 10pt
\font\ninerm=cmr9

\font\eightrm=cmr8

\newtheorem{proposition}{Proposition}

\def\qed{\hbox{${\vcenter{\vbox{                        
   \hrule height 0.4pt\hbox{\vrule width 0.4pt height 6pt
   \kern5pt\vrule width 0.4pt}\hrule height 0.4pt}}}$}}

\renewcommand{\thefootnote}{\fnsymbol{footnote}}        

\def\theequation{\thesectionc.\arabic{equation}}        

\newcommand{\di}{{\rm Div}\:}

\newcommand\R{{\Bbb R} }

\newcommand\za{{\bf a}}
\newcommand\zb{{\bf b}}
\newcommand\zd{{\bf d}}
\newcommand\zA{{\bf A }}
\newcommand\zB{{\bf B }}
\newcommand\zS{{\bf S}}
\newcommand\zF{{\bf F}}
\newcommand\zv{{\bf v }}
\newcommand\zw{{\bf w }}
\newcommand\zx{{\bf x}}

\newcommand\zp{{\bf p}}
\newcommand\zq{{\bf q}}

\newcommand\zM{{\bf M}}
\newcommand\zN{{\bf N}}
\newcommand\zE{{\bf E}}
\newcommand\zV{{\bf V}}
\newcommand\zc{{\bf c}}
\newcommand\ze{{\bf e}}
\newcommand\zu{{\bf u}}

\newcommand\0{{\bf 0}}
\newcommand\1{{\bf 1}}
\newcommand\zz{{\bf z}}
\newcommand\zZ{{\bf Z}}

\def\Mfil{{\Bbb M}}
\def\Sfil{{\Bbb S}}

\begin{document}

\runninghead{ Representation of Energy and Momentum
- October 29, 1998}
{P.~Podio-Guidugli, S.~Sellers, \& G.~Vergara Caffarelli}

 \normalsize\textlineskip
 \thispagestyle{empty}
 \setcounter{page}{1}

\copyrightheading{}                     

\vspace*{0.88truein}

\fpage{1}

\centerline{\bf ON THE REPRESENTATION OF }
\vspace*{0.070truein}
\centerline{\bf ENERGY AND  MOMENTUM IN ELASTICITY}
\vspace*{0.37truein}


      \centerline{\footnotesize P.~PODIO-GUIDUGLI}  \vspace*{0.015truein}
\centerline{\footnotesize\it Dipartimento di Ingegneria Civile}
\baselineskip=10pt
\centerline{\footnotesize\it   Universit\`a di Roma ``Tor Vergata" }
\baselineskip=10pt
\centerline{\footnotesize\it   I-00133 Roma, Italy}
\baselineskip=10pt
\vspace*{10pt}
\centerline{\footnotesize S.~SELLERS}
\vspace*{0.015truein}
\centerline{\footnotesize\it School of Mathematics}
\baselineskip=10pt
\centerline{\footnotesize\it University of East Anglia}
\baselineskip=10pt
\centerline{\footnotesize\it   Norwich NR4 7TJ, United Kingdom}
\baselineskip=10pt
\vspace*{10pt}
\centerline{\footnotesize G.~VERGARA CAFFARELLI}
\vspace*{0.015truein}
\centerline{\footnotesize\it Dipartimento di Metodi e Modelli Matematici}
\baselineskip=10pt
\centerline{\footnotesize\it Universit\`a di Roma ``La Sapienza"}
\centerline{\footnotesize\it   I-00161 Roma, Italy}
\baselineskip=10pt
\vspace*{0.225truein}
%

\vspace*{0.21truein}
\abstracts{
In order to clarify common assumptions on the form of
energy and momentum in elasticity,
a generalized conservation format is proposed for finite elasticity,
in which total energy and momentum are not specified {\it a priori}.
Velocity, stress, and total energy are assumed to depend
constitutively on deformation gradient and momentum in a manner
restricted by a dissipation principle and certain
mild invariance requirements. Under these assumptions,
representations are obtained for
energy and momentum, demonstrating that (i) the total energy splits
into separate internal and kinetic
contributions, and (ii) the momentum is linear in the velocity.
It is further shown that, if the stress response is strongly elliptic,
the classical specifications for kinetic energy and
momentum are sufficient to
give elasticity the standard format of a quasilinear hyperbolic system.
}{}{}


\vspace*{10pt}




\vspace*{1pt}\textlineskip
 \section{Introduction}
 \vspace*{-0.5pt} \noindent
 In continuum mechanics the total
 energy of a body part is generally assumed to be the sum of an internal
 energy plus  a
 kinetic energy, with the latter quadratic in the velocity; moreover,
 the momentum
 is assumed to be linear in the velocity.
 These expressions for total energy and momentum are,
 with few
 exceptions,\citeup{Ericksen,Serrina}$^{\mbox{--}}$\citeup{Serrin4,SHI1}$^{\mbox{--}}$\citeup{Shic}
 accepted without further discussion.
 One may ask, however, whether they could be derived from first
 principles. Such a derivation would perhaps allow  for alternative
 expressions, and yet lead to a plausible mechanics; or, while only
 conceptually important for well-established continuum theories, it
 would suggest a {\it modus
 operandi} in the case of continuum theories where identification
 and representation of inertial quantities is by no means obvious
 ({\it e.g.\/}, theories of liquid crystals, deformable ferromagnets,
 moving phase boundaries, crack dynamics, adhesion and peeling, \ldots).
 Recently, a derivation based upon invariance of the internal power
 and linearity of the inertial power has been
 proposed for
 classical Cauchy continua;\citeup{PPG96} the format of this
 derivation is sufficiently robust to allow for various
 generalizations.\citeup{DPG,MEGPPG1}$^{\mbox{--}}$\citeup{MEGPPG3,PPGCalcolo,PPGPeel}
 Here we follow an alternative course:
we base our approach on a presumption of
 mathematical structure, namely, that the representations for
 energy and momentum in finite elasticity be such that the resulting evolution problem
 have the form of a weakly-hyperbolic system of conservation laws,
 compatible with a dissipation inequality and certain mild invariance
 requirements.

\textheight=7.8truein
\setcounter{footnote}{0}
\renewcommand{\thefootnote}{\alph{footnote}}

 In Section 2 we first recall briefly the formulation of
 initial-value problems for general quasilinear hyperbolic systems
 of conservation laws.\citeup{Daf,Daff,Serre} Within this
 formulation, the
 standard format for nonlinear elasticity assigns the role of
 state variables to
 the deformation gradient $\zF=\partial_{x} f$ and the velocity
 $\zv=\partial_{t} f$ in a motion $f(x,t)$;
 further, the momentum is taken proportional to velocity,
 the stress $\zS=\partial_{\zF}\sigma$ is determined by the stored
 energy $\hat{\sigma}(\zF)$,\fnm{a}\fnt{a}{In elasticity,
 a purely mechanical context, stored energy replaces internal energy.} and
 the total energy $\tau$ is stored energy plus
 kinetic energy, with the latter proportional to $|\zv|^{2}$.

 We introduce next a  generalization of this conservation format for
 elasticity, in which the total energy, the stress, and the
 {\em velocity}\/    are given by general
 constitutive prescriptions $\tau=\hat\tau(\zF,\zp)$ {\it etc.\/} in terms of a pair of kinematic
 state variables, the tensor $\zF$ and the
 vector $\zp$, to be interpreted as deformation gradient and
 momentum but initially unrelated to motion. As
 a criterium of constitutive admissibility for $\hat\tau(\zF,\zp)$,
 $\hat\zS(\zF,\zp)$,  and $\hat\zv(\zF,\zp)$, we postulate that the
 purely mechanical {\em dissipation inequality}\/
$$
\dot{\hat\tau}(\zF,\zp)\;\leq\;\hat\zS(\zF,\zp)\cdot\dot\zF+\hat\zv(\zF,\zp)
\cdot\dot\zp
$$
be identically satisfied in all admissible smooth processes consistent
with the {\em conservation laws}\/
\begin{equation}
\dot{\zF}(x,t)-\di (\zv(x,t)\otimes\1)=\0, \qquad
 \dot{\zp}(x,t)-\di \zS(x,t)=\0.
\end{equation}
In this formulation of evolutionary elasticity, momentum replaces
velocity as a kinematic state variable. No relation of momentum to
mass and velocity is postulated; in particular, there is no need to
assume that the referential mass density be constant in order to
apply the general conservation format to elasticity.  The format we
propose requires that quantities describing boundary fluxes be the
subject of constitutive prescriptions; we then introduce such
prescriptions for both stress ($\equiv$ boundary flux of momentum)
and velocity ($\equiv$ boundary flux of deformation gradient).

The purpose of our generalized formulation of elasticity is to
determine the constitutive specifications for energy, momentum, and
stress that are  compatible with both the dissipation inequality
and the conservation laws. Section 3 provides these results. To
obtain them, we require arbitrary local continuations
$(\dot\zF,\dot\zp)$ of a kinematic process through $(\zF,\zp)$;
since we do not include source terms in the conservation laws, we
can only count on initial conditions. Two technical conditions,
together akin to hyperbolicity but weaker, are to be satisfied by
the constitutive mappings for velocity and stress. The first
condition is
\begin{itemize}
\item  {\em normality}\/
of $\hat\zv(\zF,\zp)$ with respect to $\zp$ for each fixed $\zF$.
\end{itemize}
A normal $\hat\zv(\zF,\zp)$ has a local inverse $\hat\zp(\zF,\zv)$
for each fixed $\zF$; it follows that the stress mapping
$\tilde\zS(\zF,\zv)=\hat\zS(\zF,\hat\zp(\zF,\zv))$ is well-defined.
The second condition is
\begin{itemize}
\item {\em ellipticity}\/ of
$\tilde\zS(\zF,\zv)$  with respect to $\zF$ for each fixed $\zv$.
\end{itemize}
Under these assumptions, arbitrariness in local
continuation is guaranteed, and thermodynamically admissible
constitutive mappings are shown to satisfy
\begin{equation}
\hat\zv(\zF,\zp)=\partial_{\zp}\hat\tau(\zF,\zp),\qquad
\hat\zS(\zF,\zp)=\partial_{\zF}\hat\tau(\zF,\zp).
\label{entropy_restrictions}
\end{equation}
These relations are necessary precursors to the representations
for velocity and energy, which are obtained under two additional constitutive
assumptions:
\begin{itemize}
\item{\em Galilean variance}\/ of velocity in a
translational change in observer,
\item{\em parity}\/ of total energy with respect to momentum.
\end{itemize}
These lead to the representations:
\begin{equation}\label{result}
\hat\zv(\zF,\zp)=\zV\zp,\qquad
\hat\tau(\zF,\zp)=\hat\kappa(\zp)+\hat\sigma(\zF),
 \qquad\hat\kappa(\zp)={1\over 2}\,\zp\cdot\zV\zp,
\end{equation}
with $\zV$ a symmetric, invertible, {\it time-independent} tensor.
We conclude that, for elasticity to have the mathematical structure
of a system of conservation laws compatible with a dissipation
inequality, momentum and kinetic energy {\em must}\/ have the
representations
\begin{equation}
\zp=\zM\zv,\qquad\kappa={1\over 2}\zv\cdot\zM\zv,
\label{mainresult}
\end{equation}
where the symmetric {\em mass-density tensor}\/ $\zM$ is the
inverse of $\zV$; moreover, since $\zM$ inherits time-independence
from $\zV$, mass {\em must} be conserved. These representations are
similar to the classical ones,\fnm{b}\fnt{b}{Namely, the
representations
$$\zp=\rho\,\zv
\qquad \mbox{and} \qquad\kappa={1\over 2}\,\rho\,\zv\!\cdot\!\zv\,,$$
with $\rho$ the referential mass density. Classically, one also
tacitly {\it assumes} the additive splitting $(\ref{result})_{2}$
of the total energy, as well as that the stress is the derivative
of the stored energy, a consequence of
$(\ref{entropy_restrictions})_{2}$ and $(\ref{result})_{2}$.} but
slightly more general due to the tensorial nature of mass density.

Relations $(\ref{result})_{2}$ and (\ref{mainresult})---the main
result of this paper---are the same arrived at when the
evolutionary problem for general Cauchy continua is put in the
invariance format introduced in Ref.$\,13$. Remarkably, the
assumptions implying (\ref{mainresult}) are compatible with {\em
hyperbolicity}\/ of the system (1.1) of conservation laws if the
mass-density tensor has the classical form $\zM=\rho\1$ (with
$\rho>0$) and  the stress mapping is strongly elliptic.


\vskip .8cm
\section{Elasticity in Conservation  Format}
\setcounter{equation}{0}
\subsection{ General hyperbolic systems with involutions}
\noindent
The general format of the initial-value
problems for quasilinear hyperbolic
systems of conservation laws
is\citeup{Daf,Daff,Serre}
\begin{eqnarray}
\partial_{t}U(x,t)+\sum_{\alpha=1}^{m}\partial_{\alpha}G^{\alpha}(U(x,t))&=&0,
\qquad\; (x,t)\in\R^{m}\times(0,\infty),\label{hyper1}\\
U(x,0)&=&U_{o}(x),\qquad x\in\R^{m}.
\label{hyper2}
\end{eqnarray}
Here $\partial_{t}$ stands for $\partial /\partial t$ and
$\partial_{\alpha}$ for $\partial /\partial x_{\alpha}$; $U$, the
{\em state vector}, takes values in a subset $\cal O$
of $\R^{n}$. The $m$ constitutive maps $G^{\alpha}$ from $\cal
O$ into $\R^{n}$ are supposed to be smooth and such as to satisfy the
following {\em hyperbolicity condition\/}:

\begin{itemize}
\item[({\sf H})] for each fixed $U$
and for each unit vector $\zw$ in $\R^{m}$, the $(n\times n)$-matrix
\begin{equation}
\sum_{\alpha=1}^{m}w_{\alpha}\,\partial_{U}G^{\alpha}\label{hyper}
\end{equation}
has real proper numbers and $n$ linearly independent proper vectors.
\end{itemize}

Generally, the systems of
conservation laws of interest for applications have an associated
{\em ``entropy''--``entropy flux''} pair $(\eta,\zq)$, with the smooth
mappings
$\eta(U)$ and $\zq(U)$ defined over $\cal O$ and taking values in $\R$
and $\R^{m}$, respectively. The role of this pair of mappings is to
characterize the admissible solutions to
(\ref{hyper1})--(\ref{hyper2}) as those satisfying the
generalized {\it ``entropy condition''}
\begin{equation}
\partial_{t}\eta(U)+\sum_{\alpha=1}^{m}
\partial_{\alpha}q^{\alpha}(U)\;\leq\;0.
\label{entro}
\end{equation}

Often the conservation system (\ref{hyper1}) has additional
geometrical structure embodied in a system of $k$ {\em
``involutions},''\cite{Daff} that is, of $k$ linear differential
equations
\begin{equation}
\sum_{\alpha=1}^{m}A_{\alpha}\,\partial_{\alpha}U=0,\qquad
A_{\alpha}= \mbox{ a constant}\,\;(k\times n)\mbox{-matrix},
\label{inv}
\end{equation}
to be satisfied by the solutions of (\ref{hyper1})--(\ref{hyper2})
so long as they are satisfied by the initial data (\ref{hyper2}).
In the presence of involutions the hyperbolicity condition
generally needs to be reformulated; in particular, the number of
linearly independent proper vectors may be less than $n$.%


\subsection{ Finite elasticity: the standard format}
\noindent
The system of finite elasticity in the referential formulation
is a standard example of a hyperbolic system with involutions (provided
the mass density is constant). To show
that the general format just described applies, consider
the elasticity system
written as \citeup{Daff,Serre}
 \begin{eqnarray}
\dot{\zF}-\nabla \zv  &=&\0, \qquad
\hbox{({\em compatibility condition})}\label{compatibility}\\
 \dot{\zv}-\di \zS    &=&\0,
\qquad \hbox{({\em momentum balance})}
\label{balance}
 \end{eqnarray}
where, for convenience, the referential mass density is presumed to
have unit value (whence the coincidence of momentum with velocity
in (\ref{balance})), and where $\zS$ is interpreted as the Piola
stress tensor.\fnm{c}\fnt{c}{As
 is common in continuum physics, we have used a superposed dot in
 place of $\partial_{t}$ and the nabla symbol in place of
 $\partial_{x}$.}

 The state vector $U$ is the pair
$(\zF,\zv)$ of a tensor $\zF$ and a vector $\zv$. To insure that these
can be interpreted, respectively, as
the deformation gradient and the velocity in a motion, that is, to
insure that
there is a motion $f(x,t)$ such that $\zF$, $\zv$, and $f$ satisfy
 \begin{equation}
 \zF(x,t)=\nabla f(x,t)\qquad{\rm and}\qquad \zv(x,t)=\partial_{t} f(x,t),
 \label{grad}
 \end{equation}
it is required that $\zF$ obey the involution following from
$(\ref{grad})_{1}$, namely,
\begin{equation}
\bigl(\nabla(\zF\za)\bigr)\zb=\bigl(\nabla(\zF\zb)\bigr)\za
\label{invol}
\end{equation}
for all constant vectors $\za$, $\zb$
({cf.}~(\ref{inv})).\fnm{d}\fnt{d}{As is well-known, given the pair
$(\zF,\zv)$, the kinematical compatibility conditions
(\ref{compatibility}) and (\ref{invol}) are sufficient for the
existence of a local solution $f$ of the system (\ref{grad}).
\hfill\linebreak[4]
Here and henceforth, unless otherwise
specified a {\it vector} is meant to be an element of a {\it
three-dimensional\/} inner-product vector space $\cal V$, and a
{\it tensor} is a linear transformation of $\cal V$ into itself.
Moreover, whenever needed or simply convenient, $\cal V$ will be thought of
as being endowed with an orthonormal basis $\{{\bf c}_{\alpha},\,
\alpha=1,2,3\}$. All components of vectors, tensors, {\it etc.\/} will
be tacitly taken with respect to this basis and related
constructs.}

As to the ``entropy''--``entropy flux'' pair, one chooses
\begin{eqnarray}
\eta(\zF,\zv)&=&\kappa(\zv)+\sigma(\zF),\\
\zq(\zF,\zv)&=&-\zS^{\top}\!(\zF)\,\zv,
\end{eqnarray}
with
\begin{equation}
\kappa(\zv)={1\over 2}\,\zv\!\cdot\!\zv,
\label{kin}
\end{equation}
the {\em kinetic energy}, $\sigma(\zF)$ the
{\em stored energy}, and
\begin{equation}
\zS(\zF)=\partial_{\zF}\sigma(\zF).
\end{equation}
Thus the ``entropy'' in this case is  the total energy
$(\kappa+\sigma)$ per unit volume, the ``entropy flux'' is the flux
necessary to balance the total energy, and the generalized
``entropy condition'' (\ref {entro}) takes  the simple form
\begin{equation}
\partial_{t}(\kappa+\sigma)-\di (\zS^{\top}\zv)=0.
\end{equation}
 Remarkably, this relation
is consistent with the Second Law of thermodynamics, when the latter
is formulated in the form of the Clausius-Duhem inequality.
%
%

\subsection{ Finite elasticity: a generalized format}
\noindent Consider now the following generalization of the
standard conservation format for finite elasticity.

For the state
vector $U=(\zF,\zp)$, and for $\1$ the identity tensor, let the
initial-value problem (\ref{hyper1})--(\ref{hyper2}) take the form
\begin{eqnarray}
\dot{\zF}(x,t)-\di \bigl(\zv(x,t)\otimes\1\bigr)&=&\0, \label{comp2}\\
 \dot{\zp}(x,t)-\di \zS(x,t)&=&\0, \label{bal2}\\
 \zF(x,0)\ =\ \zF_{o}(x),\qquad \zp(x,0)&=&\zp_{o}(x).\label{iniz}
 \end{eqnarray}
Next, as the generalized format suggests,
let the conservation laws (\ref{comp2})
and (\ref{bal2}) for the state variables ``deformation gradient'' $\zF$
and ``momentum'' $\zp$  be
supplemented by {\em constitutive
prescriptions for velocity and stress}\/ as functions of state:
\begin{eqnarray}
\zv&=&\hat\zv(\zF,\zp),\label{velocity_const_rel}\\
\zS&=&\hat{\zS}(\zF,\zp).\label{stress_const_rel}
\end{eqnarray}
Finally, let the generality of these constitutive prescriptions and
of the additional prescription for the {\em total energy}:
\begin{equation}
\tau=\hat\tau(\zF,\zp),\label{energy_const_rel}
\end{equation}
be restricted by requiring consistency with the Second Law, under form of the
dissipation inequality
\begin{equation}
\dot{\hat\tau}(\zF,\zp)\;\leq\;\hat\zS(\zF,\zp)\cdot\dot\zF+\hat\zv(\zF,\zp)
\cdot\dot\zp,
\label{dissi}
\end{equation}
to be identically satisfied in all admissible smooth
processes.\fnm{e}\fnt{e}{In all constitutive mappings we leave
tacit a possible explicit dependence on the space variable, but
exclude any explicit time dependence.}

As to our present formulation of finite elasticity, some comments
are in order. Firstly, the state variables for system
(\ref{comp2})--(\ref{bal2}) are a tensor $\zF$ and a vector $\zp$.
Formally,  neither $\zF$ need be the gradient of a deformation
(this geometric compatibility problem may be solved once a solution
of the initial-value problem (\ref{comp2})--(\ref{iniz}) has been
found, but at this stage we do not require that $\zF$ satisfy the
involution (\ref{invol})) nor that $\zp$ need have any {\em a
priori}\/ specified relation with mass density and velocity. The
compatibility condition (\ref{compatibility}) is generalized to the
evolution equation (\ref{comp2}), which we interpret as the balance
of $\zF$ by way of the boundary-flux tensor $\zv\otimes\1$.
Similarly, the momentum balance (\ref{balance}) is generalized to
(\ref{bal2}), the balance of $\zp$ by way of the boundary-flux
tensor $\zS$. Less formally, we regard (\ref{bal2}) as a force
balance, with $-\dot\zp$ the inertial force (the only external
force considered here) and $\di \zS$ the internal force per unit
referential volume.

Secondly, just as
the momentum is not given any {\em a priori}\/ representation, the other
inertial manifestation---kinetic energy---is not given any {\em a priori}\/
representation as well. In fact, neither the total energy is
split into a kinetic part
and an internal part nor the Piola stress is introduced by
differentiation of the latter.

Our next step is to introduce and exploit a set of reasonable hypotheses
allowing to conclude that slight generalizations of the
 classical constitutive
specifications for energy, momentum, and stress, are the only
specifications compatible with both the conservation laws
(\ref{comp2}), (\ref{bal2}) and the dissipation inequality
(\ref{dissi}).

%
%
\section{Representations of Energy and Momentum}
\setcounter{equation}{0}
\vskip -.2cm
\subsection{Normality and ellipticity. Thermodynamical
admissibility}
\noindent As is customary in
constitutive theories of continuous media after a famous paper by
Coleman \& Noll,\cite{{CN}} the dissipation inequality
restricts the choice of admissible representations of the
constitutive mappings. These restrictions are derivable under the
crucial assumption that any accessible kinematic state $(\zF,\zp)$ can be
reached by a path that has an arbitrary local continuation
$(\dot\zF,\dot\zp)$. Such indispensable arbitrariness is usually
taken to be granted by the presence in the basic balance equations
of source terms regarded as
controls at our disposal. Since sources are here set to
null, we revert to the other piece of data that in principle can
be made to vary arbitrarily, that is, initial conditions.\fnm{f}\fnt{f}{This
idea is not new,\citeup{Ingo,Mega,Megb,Liu} although we know neither of any
previous use in the context of
elasticity nor of anything more than a generic claim of local
existence for the initial-value problem under examination.} $\,$Our first proposition establishes that such an
indispensable arbitrariness is indeed granted under
reasonable conditions on the balance equations and the
constitutive mappings for velocity and stress.

The constitutive mappings $\hat\zv$ and $\hat\zS$ are taken to be
continuously differentiable and to satisfy two conditions, one
each. The first condition applies to $\hat\zv$:

\begin{itemize}
\item[({\sf N})](Normality) For each $\zF$ fixed, the map
$\zp\mapsto \hat\zv(\zF,\zp)$ is {\it normal}, {\it i.e.\/}, the
matrix $\zN$, defined by
$$
N_{ih}(\zF,\zp)=\partial_{p_{h}}\hat v_{i}(\zF,\zp),
$$
is invertible for all $\zp$.
\end{itemize}

\noindent This condition formalizes the expectation that momentum and
velocity be locally in one-to-one correspondence, in the sense that
the map $(\zF,\zv)\mapsto(\zF,\zp=\hat \zp(\zF,\zv))$ is
well-defined, with
$$
\hat\zv(\zF,\hat\zp(\zF,\zv))=\zv.
$$
Let now
\begin{equation}\label{tildeS}
\tilde\zS(\zF,\zv):=\hat\zS(\zF,\hat\zp(\zF,\zv)).
\end{equation}
The second condition, rather than directly to $\hat \zS(\zF,\zp)$,
applies to $\tilde\zS(\zF,\zv)$:
\begin{itemize}
\item[({\sf E})](Ellipticity) For each $\zv$ fixed, the map
$\zF\mapsto \tilde\zS(\zF,\zv)$ is {\it elliptic} in the sense of
Petrovsky, {\it i.e.\/}, the matrix $\zE$, defined by
$$
E_{ih}(\zF,\zv;\za)=\partial_{F_{hk}}\tilde
S_{ij}(\zF,\zv)\,a_{j}a_{k}
\quad{\rm for}\;\,{\rm each}\;\,{\rm vector}\;\,\za,
$$
is invertible for all $\zF$.
\end{itemize}
Normality of $\hat\zv(\zF,\cdot)$ and ellipticity of
$\tilde\zS(\cdot,\zv)$ do not imply hyperbolicity of the system
(\ref{comp2})--(\ref{bal2}). As discussed later in this section,
some strengthening of these assumptions is needed to obtain
hyberbolicity. For the time being, we consider the following {\it
initial-value problem\/}:

\begin{itemize}
\item[]For each ${T}>0$ and $x\in\R^3$ fixed, find $t\mapsto
(\zF(x,t),\zp(x,t))$ such that
\begin{eqnarray}
\dot{\zF}(x,t)&=&\nabla \zv(x,t), \label{comp3}\\
 \dot{\zp}(x,t)&=&\di \tilde\zS(\zF(x,t),\zv(x,t)), \label{bal3}
\end{eqnarray}
for $t\in (0,{T})$ and, moreover,
\begin{eqnarray}
 \zF(x,0)&=&\zA + \za\!\cdot\! (\zx-\zx_{o})(\zb\otimes\za),\label{iniz1}\\
 \zv(x,0)&=&\zB(\zx-\zx_{o})+\zc.\label{iniz2}
\end{eqnarray}
\end{itemize}

Here $x_{o}$ is a given point of $\R^3$, $\zA$ and $\zB$ are two
given matrices, and $\za$, $\zb$, and $\zc$ three given vectors.
This problem is obtained by supplementing
(\ref{comp2})--(\ref{bal2}) by the constitutive relation
(\ref{tildeS}) and the initial conditions
(\ref{iniz1})--(\ref{iniz2}), which involve the state variables
$(\zF,\zv)$; note that, due to ({\sf N}),
(\ref{iniz1})--(\ref{iniz2}) induce a unique pair of initial
conditions for the state variables $(\zF,\zp)$.
\vspace*{12pt}
\begin{proposition} [Arbitrariness of initial time-rates of the state variables]
\tenrm Assume that the constitutive
mappings $\hat\zv$ and $\tilde\zS$ are, respectively, normal and
elliptic. Moreover, assume that, for each fixed triplet
$(\zA,\za,\zc)$, the initial-value problem
(\ref{comp3})--(\ref{iniz2}) has a classical solution up to time
${T}$ for all choices of $\zB$, $\zb$, and $x_{o}$. Then
 the initial time-rate $\bigl(\dot\zF(x_{o},0),\dot\zp(x_{o},0)\bigr)$
at $x_{o}$ of the state vector can be assigned arbitrary values.
\end{proposition}
\vspace*{12pt}
\noindent {\bf Proof.} Note firstly that the evolution equations
(\ref{comp3})--(\ref{bal3}) can be written as
\begin{eqnarray}
\dot F_{ij}&=&\partial_{j}v_{i},\\
\dot p_{i}&=&\partial_{F_{hk}}\tilde S_{ij}(\zF,\zv)\,
\partial_{j}F_{hk}+
\partial_{v_{k}}\tilde S_{ij}(\zF,\zv)\,\partial_{j}v_{k}.
\end{eqnarray}
Moreover,  the initial conditions (\ref{iniz1})--(\ref{iniz2})
yield
\begin{eqnarray}
(\zF(x_{o},0),\zv(x_{o},0))&=&(\zA,\zc),\\
(\partial_{j}F_{hk}(x_{o},0),\partial_{j}v_{k}(x_{o},0))&=&(b_{h}a_{k}a_{j},B_{kj}).
\end{eqnarray}
Then, since the initial-value problem (\ref{comp3})--(\ref{iniz2})
has classical solutions, with the notations introduced in the
statement of condition ({\sf E}) we have that
\begin{eqnarray}
\dot F_{ij}(x_{o},0)&=&B_{ij},\\
\dot p_{i}(x_{o},0)&=&\partial_{v_{k}}\tilde S_{ij}(\zA,\zc) \,
B_{kj}+E_{ih}(\zA,\zc;\za)\,b_{h}.
\end{eqnarray}
When {\bf B} and {\bf b} are arbitrarily varied, the invertibility
of $\zE(\zA,\zc;\za)$ guarantees that the second term on the right
of the last relation and, hence, $\dot\zp(x_{o},0)$ take arbitrary
values, not only $\dot\zF(x_{o},0)$.\qquad\qed
\vspace*{12pt}
\begin{proposition}[Thermodynamically admisssible constitutive
relations] \hfill\break
\tenrm
Under the assumptions of Proposition 1, necessary and sufficient
conditions for the dissipation inequality (\ref{dissi}) to be
satisfied in all admissible smooth processes of the materials
described by the constitutive relations (2.18)--(2.19) are
\begin{eqnarray}
\hat\zv(\zF,\zp)&=&\partial_{\zp}\hat\tau(\zF,\zp),\\
\hat\zS(\zF,\zp)&=&\partial_{\zF}\hat\tau(\zF,\zp).
\label{thermo_admiss}
\end{eqnarray}
Moreover, if $\hat\tau$ is twice continuosly differentiable, then
\begin{equation}
\partial_{\zp}\hat\zS=\partial_{\zF}\hat\zv.
\end{equation}
\end{proposition}
\vspace*{12pt}
\noindent{\bf Proof.} From (2.21) and by the chain rule,
\begin{equation} (\partial_\zF \hat\tau-\hat\zS)\! \cdot \!\dot{\zF}+
(\partial_\zp \hat\tau -\hat\zv)\!\cdot\! \dot{\zp}\; \le \;0.
\nonumber
\end{equation}
Sufficiency of (3.11)--(3.12) follows by substitution.
Since $(\dot{\zF},\dot{\zp})$ can be
chosen arbitrarily, at least initially (Proposition 1), we obtain
necessity. \qquad \qed
%
%
%

\subsection{ Galilean variance and parity. Representation results}
\noindent It is common in mechanics to impose some type of invariance
requirements on the constitutive relations. Here we consider a
notion of {\it translational change in observer} appropriate to our
present choice of the state vector. Since neither $\zF$ nor $\zp$
are presumed to have any relation to the motion, we cannot deduce
their variance laws from the standard notion of observer change.
Instead, we say that the pairs $(\zF,\zp)$ and $(\zF^{*},\zp^{*})$
are related by a translational change in observer whenever there is
a vector $\zd$, independent of time, such that
\begin{eqnarray}
\zF\mapsto \zF^{*}&=&\zF,\nonumber\\
\zp\mapsto\zp^{*}&=&\zp+\zd.
\end{eqnarray}
Thus, while $\zF$ is invariant, $\zp$ is not; $\zd$ measures the
``invariance defect'' of the latter. Note that when the momentum $\zp$ is given the
classical representation $\zp=\rho\zv$, this notion reduces to the
standard one, provided one chooses $\zd=\rho\ze$, with $\ze$ a
constant vector. As to the behavior in a translational observer
change of the constitutive mapping delivering the velocity, we
introduce the following assumption.

\begin{itemize}
\item[({\sf G})](Galilean Variance of Velocity)
In a translational change in
observer,
\begin{equation}
\hat\zv(\zF,\zp+\zd)-\hat\zv(\zF,\zp)=\hat\zc_{\zv}(\zd)
\qquad \mbox{for each vector}\;\,\zd,
\end{equation}
where the mapping $\hat\zc_{\zv}$ specifies the invariance defect
of the velocity mapping $\hat\zv$.
\end{itemize}
Our next assumption specifies the parity of total energy with respect to
the second state variable, the momentum $\zp$:

\begin{itemize}
\item[({\sf P})](Parity of Total Energy with Respect to Momentum)
For each $\zF$ fixed, the total energy mapping is even in its second
argument, namely,
\begin{equation}
\hat \tau(\zF,\zp)=\hat \tau(\zF,-\zp).
\end{equation}

\end{itemize}
Note that, due to (3.11), parity of total energy renders the
velocity odd:
\begin{equation}
\hat \zv(\zF,\zp)=-\hat \zv(\zF,-\zp).
\end{equation}

The next two propositions state our main representation results.
\vskip .2cm
\begin{proposition}[Constitutive representation of velocity]
\tenrm Let the constitutive mappings $\hat\zv$ and $\hat\tau$
satisfy, respectively, the variance and parity requirements ({\sf
G}) and ({\sf P}). Then $\hat\zv$ has the representation
 \begin{equation}
 \hat\zv(\zF,\zp)=\zV\zp, \label{rep_velocity}
 \end{equation}
 where the symmetric and invertible tensor $\zV$ depends at most on $x$.
 \end{proposition}
\noindent {\bf Proof.} By assumption ({\sf G}),
\begin{eqnarray}
\partial_{\zF}\hat\zv(\zF,\zp+\zd)&=&\partial_{\zF}\hat\zv(\zF,\zp),\\
\partial_{\zp}\hat\zv(\zF,\zp+\zd)&=&\partial_{\zp}\hat\zv(\zF,\zp),
\end{eqnarray}
so that, both $\partial_{\zF}\hat\zv$ and $\partial_{\zp}\hat\zv$ do
not depend on $\zp$. Relation (3.21) then yields the preliminary
representation
\begin{equation}
\hat\zv(\zF,\zp)=\tilde\zV(\zF)\zp+\tilde\zv(\zF),\nonumber
\end{equation}
which, by consequence (3.18) of assumption ({\sf P}), reduces to
\begin{equation}
\hat\zv(\zF,\zp)=\tilde\zV(\zF)\zp.
\end{equation}
Consistency with (3.20) requires that $\tilde\zV(\zF)\equiv\zV$, with $\zV$
depending at most on $x$. Moreover, the normality assumption ({\sf N})
guarantees that the tensor {\bf V} is invertible. Finally, it
follows from (3.11) and (3.23) that {\bf V} is symmetric. \qquad\qed
\vspace*{12pt}
\begin{proposition}[Constitutive representation of
energy]
\tenrm If Propositions (1)--(3) hold, then the total energy
$\hat\tau(\zF,\zp)$ has the representation
 \begin{equation}
 \hat\tau(\zF,\zp)=\hat\kappa(\zp)+\hat\sigma(\zF),
 \qquad\hat\kappa(\zp)={1\over 2}\,\zp\cdot\zV\zp,
\label{rep_energy}
 \end{equation}
 where $\hat\kappa(\zp)$ is interpreted as the
 {\it kinetic energy} and
$\hat\sigma(\zF)$ as the {\it stored energy}.
\end{proposition}
\vspace*{12pt}
\noindent {\bf Proof.} By (3.12), (3.13), and (3.19), we have that
\begin{equation}
\partial^{(2)}_{\zF\zp}\hat\tau(\zF,\zp)=\0,
\end{equation}
which yields (\ref{rep_energy})$_1$.
The representation (\ref{rep_energy})$_2$ then follows from (3.11) and again
(3.19).\qquad\qed
\vspace*{12pt}

\noindent Because of the presence of the tensor $\zV$,
the expressions for $\hat\zv$
and $\hat\kappa$ are slightly more general than the usual ones.
 As expected, the stored energy determines the stress,
since (\ref{thermo_admiss}) and (\ref{rep_energy})$_1$ immediately yield
\begin{equation}
\hat\zS(\zF)=\partial_{\zF}\hat\sigma(\zF). \label{rep_stress}
\end{equation}
We note that, since (the velocity mapping is independent of $\zF$
and) the stress mapping is independent of $\zp$, the ellipticity
condition ({\sf E}) can be phrased in terms of $\hat\zS(\zF,\zp)$.


\subsection{ Constitutive representations and hyperbolicity}
\noindent
We have proposed a set of reasoned assumptions under which the
general constitutive relations (\ref{velocity_const_rel}),
(\ref{stress_const_rel}), and (\ref{energy_const_rel})  take,
respectively, the forms (\ref{rep_velocity}), (\ref{rep_stress}),
and (\ref{rep_energy}). The latter group of relations implies
hyperbolicity of the elasticity system, provided further
assumptions are introduced. Such hyperbolicity---which, as is
well-known, cannot be strict---is described precisely in our last
proposition.

As a premise, we note that, for
\begin{equation}\label{ellipt}
\Sfil=\partial_{\zF}\hat\zS=\partial_{\zF\zF}^{(2)}\hat\sigma\,,
\end{equation}
the ellipticity matrix can be written as a linear transformation of
vectors:
\begin{equation}
\zE(\zw)\zu=(\Sfil[\zu\otimes\zw])\zw,\qquad\zw,\zu\in{\cal V},\quad
|\zw|=1.
\end{equation}
Due to (\ref{ellipt}), {\it strong ellipticity} of the stress
mapping $\hat\zS$ (that is, positivity of $\zE(\zw)$ for each unit
vector $\zw$) is equivalent to {\it strict {\rm (}rank 1\/{\rm
)}-convexity\/} of the stored-energy mapping $\hat\sigma$.
\vspace*{12pt}
\begin{proposition}[Hyperbolicity of Elasticity in Conservation
Format]
\tenrm Let \linebreak[4]
$\zV=\rho^{-1}\1$, where $\rho>0$ is interpreted as the referential
{\it mass density}. Moreover, let the matrix ${\bf E}(\zw)$ be
positive for each unit vector {\bf w}. Then, for each $\zw$, the
$(12\!\times\! 12)$-matrix that now corresponds to the general
hyperbolicity matrix (\ref{hyper}) has zero as proper value of
geometric multiplicity six, and has, moreover, exactly three
independent proper vectors corresponding to non-null proper numbers
and giving the possible directions of wave
propagation.\fnm{g}\fnt{g}{$\;\,$The occurrence of zero as a proper
number with high multiplicity is due to the involutions
(\ref{invol}), which render the elasticity system not strictly
hyperbolic ({cf.}~Dafermos\citeup{Daff}).}
\end{proposition}
\vspace*{12pt}
\noindent {\bf Proof.} Under the assumptions, for each fixed $\zw$,
the problem we study can be given the form
\begin{equation}
\Mfil \,(\zw) \,V=\lambda\, V,\label{eq:m}
\end{equation}
where the entries of the $(4\times 4)$-blockmatrix $\,\Mfil\,$ that satisfies
(\ref{hyper}) are the following tensors:
\begin{equation}\qquad\left\delimiter0
\begin{array}{c@{\:=\:}l}
\Mfil_{\alpha\beta} & \Mfil_{44}\;=\;\0,\phantom{\displaystyle\frac{1}{2}}    \\
\Mfil_{\alpha 4} & -\rho^{-1}{\bf w}\otimes{\bf c}_{\alpha},\phantom{\displaystyle\frac{1}{2}}  \\
\Mfil_{4\alpha}{\bf u} & -(\Sfil[{\bf u}\otimes{\bf c}_{\alpha}]){\bf w}, \phantom{\displaystyle\frac{1}{2}}
\end{array} \qquad
\begin{array}{l}
(\alpha,\,\beta=1,2,3),\phantom{\displaystyle\frac{1}{2}} \\ \phantom{\displaystyle\frac{1}{2}} \\
\qquad{\bf u}\in {\cal V}, \phantom{\displaystyle\frac{1}{2}}
\end{array}\label{m} \right\}
\end{equation}
and where, given a tensor {\bf Z} and a vector {\bf z},
the entries of the 4-blockvector $V$ are the vectors
\begin{equation}
{\bf v}_{\alpha}={\bf Z}{\bf c}_{\alpha}\;\;(\alpha=1,2,3),
\qquad{\bf v}_{4}={\bf z}.
\end{equation}
With (\ref{m}) the equation (\ref{eq:m}) yields the system
\begin{eqnarray}
\Mfil_{\alpha 4}\zv_{4}&=&\lambda\zv_{\alpha},\\
\sum_{\alpha=1}^{3}\Mfil_{4\alpha }\zv_{\alpha}&=&\lambda\zv_{4},
\end{eqnarray}
or rather, more explicitly,
\begin{eqnarray}
-\rho^{-1}\zz\otimes\zw&=&\lambda\zZ, \label{eq:system1}\\
-(\Sfil[\zZ])\zw&=&\lambda\zz. \label{eq:system2}
\end{eqnarray}

For $\lambda=0$, (\ref{eq:system1}) implies that $\zz=\0$; moreover, choosing
$\zc_{3}=\zw$, we can represent $\zZ$ as
\begin{equation}
\zZ=\zv_{1}\otimes\zc_{1}+\zv_{2}\otimes\zc_{2}+\zv_{3}\otimes\zw;
\end{equation}
consequently, (\ref{eq:system2}) becomes
\begin{equation}
(\Sfil[\zv_{3}\otimes\zw])\zw=\zE(\zw)\zv_{3}=
-(\Sfil[\zv_{1}\otimes\zc_{1}+\zv_{2}\otimes\zc_{2}])\zw,
\end{equation}
whence, in view of the assumed positivity of $\zE(\zw)$, we conclude that
$\zZ$ belongs to a six-dimensional subspace of the space of all tensors.
Moreover, assuming provisionally that $\lambda \neq 0$, one
finds by inserting (\ref{eq:system1}) into ( \ref{eq:system2}) that
\begin{equation}
\zE(\zw)\zz=\mu\,\zz,\qquad\mu=\rho\,\lambda^{2}.
\end{equation}
Hence, the positivity of $\zE(\zw)$ implies that  there are {\it positive}
proper numbers $\mu$ and exactly three independent proper vectors.
\qquad\qed
\vspace*{12pt}

\noindent We remark that the involution condition
(\ref{invol})---which we do not use---implies that the proper
space associated to the proper number $\lambda=0$
is six-dimensional.\cite{Daff}
%

\nonumsection{Acknowledgments}
\noindent The work of Podio-Guidugli and Vergara Caffarelli was supported by
M.U.R.S.T. within, respectively, the Progetto Nazionale ``Termomeccanica dei Continui Classici
e dei Materiali Nuovi'' and the Progetto Nazionale ``Equazioni
Differenziali e Calcolo delle Variazioni'', whereas
 Sellers held a fellowship for
foreign mathematicians of the Consiglio Nazionale delle Ricerche.

\vskip 16pt

\nonumsection{References}
\end{document}